\documentclass[aps, 11pt, A4, nofootinbib, preprint numbers]{revtex4-1}
\usepackage{float}
\usepackage{amsmath}
\usepackage{graphicx}
\usepackage{subcaption}
\usepackage{color}
\usepackage{amssymb}
\usepackage{amsfonts}
\usepackage{natbib}
\usepackage{multirow}
\usepackage{hhline}

\newcommand{\be}{\begin{equation}} 
\newcommand{\ee}{\end{equation}} 
\newcommand{\bea}{\begin{eqnarray}} 
\newcommand{\eea}{\end{eqnarray}} 
\newcommand{\bqa}{\begin{eqnarray}}
\newcommand{\eqa}{\end{eqnarray}}
\newcommand{\bwt}{\begin{widetext}}
\newcommand{\ewt}{\end{widetext}}
\newcommand{\mb}{\mathbf}

\newcommand{\nn}{\nonumber \\}

\newcommand{\w}{\omega}

\begin{document} 
\title {Non-equilibrium electron relaxation in Graphene}
\author{Luxmi Rani}\email{luxmiphyiitr@gmail.com}
\author{Pankaj Bhalla}
\author{Navinder Singh}
\affiliation{Theoretical Physics Division, Physical Research Laboratory, Ahmedabad-380009, India.} 
\date{\today}

\begin{abstract}
We apply the powerful method of memory function formalism to investigate non-equilibrium electron relaxation in graphene. Within the premises of Two Temperature Model (TTM), explicit expressions of the imaginary part of the Memory Function or generalized Drude scattering rate ($1/\tau$) are obtained. In the DC limit and in equilibrium case where electron temperature ($T_e$) is equal to phonon temperature (T), we reproduce the known results (i.e. $ 1/\tau \propto T^4$ when $T<<\Theta_{BG}$ and $1/\tau \propto T$ when $T>>\Theta_{BG}$,  where $ \Theta_{BG}$ is the Bloch-Gr\"{u}neisen temperature). We report several new results for $1/ \tau$ where $T \neq T_e$ relevant in pump-probe spectroscopic  experiments. In the finite frequency regime we find that $1/\tau \propto \omega^2$ when $\omega<<\omega_{BG}$, and for $\omega>>\omega_{BG}$ it is $\omega$ independent and also electron temperature independent. These results can be verified in a typical pump-probe experimental setting for graphene.
\end{abstract}
\maketitle

\section{Introduction}
 Graphene is a unique two dimensional material consisting of a single atom
thick layer of carbon atoms that are closely packed in honeycomb lattice structure. In recent times, the study of electronic transport of hot carriers in graphene has created an enormous research interest in both the experimental and theoretical aspects due to the potential applications in electronic devices \cite{Novoselov,Neto,Allen,Sarma,LI,NMR,Rao, Shah}. In graphene, relaxation of hot (photoexcited) electrons has been investigated experimentally in \cite{Sarma, LI,Gabor,SW, KJT,Mak}and theoretically in \cite{Dvgaev,Kim,Low,Iglesias,Tse,Butscher,EM, Efetov,EH,Tan,Fuhrer}. In simple metals, electron relaxation dynamics is well understood and the two temperature model (TTM) is extensively used to analyze the relaxation dynamics \cite{Wong,Verburg,Majchrzak,Chen, NS,Das}. While, in graphene due to Dirac physics and peculiar band structure, hot electron relaxation is different from that metal, and a detailed theoretical study is lacking. 

In simple metals, hot electron relaxation happens via electron-phonon interactions. The mechanism of hot electron relaxation is as follows. A Femto-second laser pulse  excites the electrons from equilibrium Fermi-Dirac (FD) distribution to a non-equilibrium distribution. This non-equilibrium electron distribution internally relaxes via electron-electron interactions to a hot FD-distribution in a time scale $\tau_{ee}$. Then through electron-phonon interactions, this ``hot" FD-distribution relaxes to a state in which electron temperature becomes equal to the phonon temperature i.e., an equilibrium state. This process happens in a time scale $\tau_{e-ph}$. In simple metals the inequality $\tau_{ee}<<\tau_{e-ph}$ is true. And phonons remain in equilibrium during the whole process of relaxation (it is called the Bloch assumption\cite{NS}). This motivates the two temperature model (TTM): one temperature for electron sub-system ($T_e$) and another for the phonon sub-system (T). The electron relaxation in metals is extensively studied within TTM model using the Bloch-Boltzmann kinetic equation\cite{Majchrzak,Chen,NS,Das}. In the analysis an important energy scale is set by Debye temperature, and it turns out that when $T>>\Theta_{D}$, the relaxation rate from the Bloch-Boltzmann equation is given as $1/\tau \propto T$. In the opposite limit, i.e., $T<<\Theta_{D}$ it turns out that $1/\tau \propto T^5$.

In order to study the hot electron relaxation in graphene, several experiments like pump - probe spectroscopy and photo-emission spectroscopy has been used recently \cite{BAR, Paul, Liaros}. On the theoretical side, the hot electron relaxation has been studied in graphene using the Bloch-Boltzmann equation \cite{Dvgaev,Kim, EM}. But all these studies are restricted to the DC regime.

A detailed study of frequency and temperature dependent scattering rate in graphene has been lacking in the literature. In the present investigation, we solved this problem using the powerful method of memory function formalism\cite{Singh,GW,Kubo}. We calculate the scattering rate in various frequency and temperature limits. Our main results are ;

 In the DC case, scattering rate shows the fourth power law of both electron and phonon subsystem temperatures below the BG temperature. Above the BG temperature, scattering rate is linearly dependent on phonon temperature only. On the other hand, at higher frequency and at higher temperature, scattering rate is independent on frequency and electron temperature. It is observed that there is $\w^2$-dependence in the lower frequency regime.

  This paper is organized as follows. In section \ref{sec:theory}, we introduce the model and memory function formalism. We then compute the memory function (generalised Drude scattering rate) using the W\"olfle-G\"otze perturbative method\cite{GW}. Then various sub-cases are studied analytically. In section \ref{sec:numeric}, we present the numerical study of the general case. Finally, we summarize our results and present our conclusions.

\section{Theoretical Framework}
\label{sec:theory}

 To study the electron relaxation in graphene, we consider total Hamiltonian having three parts such as free electron ($H_{\text{e}}$), free phonon ($H_{\text{p}}$) and interacting part i.e electron-phonon ($H_{\text{ep}}$):
 \bea
 H & = & H_{\text{e}} + H_{\text{p}} + H_{\text{ep}}.
 \label{eq:H} 
 \eea
 The different parts of Hamiltonian mentioned in the above equation are defined as
\bea
H_{\text{e}} &= & \sum_{\mb{k}\sigma} \epsilon_k c^\dagger_{\mb{k}\sigma}c_{\mb{k}\sigma},\\
H_{\text{p}}& = &\sum_{q}\omega_q\left(b^\dagger_q b_q +\frac{1}{2}\right),\\
H_{\text{ep}} & = & \sum_{\mb{k},\, \mb{k'},\sigma}\left[ D(\mb{k}-\mb{k'})c^\dagger_{\mb{k}\sigma}c_{\mb{k'}\sigma}b_{\mb{k}-\mb{k'}} + \text{H.c.} \right].
 \label{eq:H1} 
\eea
Here, $c^\dagger_{\mb{k}\sigma} (c_{\mb{k}\sigma})$ and $b^\dagger_q(b_q)$ are electron and phonon creation (annihilation) operators, $\sigma$ is a spin, $\mb{k}$ and  $\mb{q}=\mb{k}-\mb{k'}$ are electron and phonon momentum respectively. $\epsilon_k=\hbar v_F|\vec{k}|$ is the linear energy dispersion term in graphene. $D(\mb{k}-\mb{k'})$ is the electron-phonon matrix element which is defined as\cite{EM, Mahan,Ziman}
\bea
D(q)&=&-i \bigg(\frac{1}{2\rho_m \w_q}\bigg)^ {1/2} D_0\times q \left[1-\bigg(\frac{q}{2k_F}\bigg)^2\right]^{1/2}.
\label{eq:e-ph-mat}
\eea
Here, $D_0$ is the deformation potential coupling constant for graphene, $ \rho_m $ is surface mass density and $k_F$ is the Fermi momentum and $\w_q$ is the phonon energy. Here, we set $\hbar=k_B=1$ throughout the calculations.

\subsection {Calculation for generalized Drude scattering rate}
Our aim is to calculate the generalized Drude scattering rate or imaginary part of the memory function. In a typical experimental set-up, reflectivity from a graphene sample is measured at various frequencies; and it is written as \cite{Mahan, Singh}:
  \bea
  R(\omega)&=&\frac{(n(\omega)-1)^2 + (k(\omega))^2}{(n(\omega)+1)^2 + (k(\omega))^2},
  \eea
     Where, 
  \bea
    n(\omega)&=&\frac{1}{\sqrt{2}}\sqrt{\sqrt{\epsilon_1^2(\omega) + \epsilon_2^2(\omega) } + \epsilon_1(\omega)},
    \eea
    \bea
    k(\omega)&=&\frac{1}{\sqrt{2}}\sqrt{\sqrt{\epsilon_1^2(\omega) + \epsilon_2^2 (\omega)} -\epsilon_1(\omega)}.
 \eea
  $\epsilon_1(\omega)$ and $\epsilon_2(\omega)$ are the real and imaginary parts of the dielectric function which are related to real and imaginary parts of the conductivity ($\sigma(\omega)$). Thus, from the reflectivity data, frequency dependent conductivity can be obtained \cite{Singh}. From conductivity data, by Kramers-Kronig (KK) analysis, real and imaginary parts of the memory function are obtained as the conductivity can be written as \cite{GW}:
\bea
\sigma(\omega)&=&-i\frac{1}{\omega + M(\omega)}.
\eea
For the calculation of generalized Drude scattering, we use the G\"otze-W\"olfle formalism \cite{Das,Singh, LR}. In this formalism, memory function is expressed as
\bea
M(z,T,T_e)&=&\frac{z \chi(z)}{\chi_0-\chi(z)} \simeq \frac{z\chi(z)}{\chi_0}\bigg(1 + \frac{\chi(z)}{\chi_0} +....\bigg) \nn
 &&\simeq \frac{z\chi(z)}{\chi_0},
\label{eq:M1}
\eea
where, $\chi_0$ represents the static limit of correlation function (i.e. $ \chi_0 = Ne/m $) and $\chi(z)$ is the Fourier transform of the current-current correlation function:
 \bea 
 \chi(z)= i\int_0^\infty e^{izt}\langle [j_1, j_1]\rangle dt.
 \eea 
Here, $j_1=\Sigma_{k\sigma}(\vec{k}.\hat{n})c^\dagger_{\mb{k}\sigma}c_{\mb{k}\sigma}$ is the current density. $\hat{n}$ is the unit vector along the direction of current. Using the equation of motion (EOM) method \cite{GW, Singh} it can be shown that
 \bea
 M(z, T, T_e)&=&\frac{\langle\langle [j_1, H];[j_1, H]\rangle\rangle_{z=0}-\langle\langle [j_1, H];[j_1, H]\rangle\rangle_z}{z\chi_0}.
 \label{eq:M}
 \eea
 Substituting equation (\ref{eq:H}) and the definition of current density operator into the above equation and on simplifying \footnote{The current density operator commutes with the non-interacting parts of the Hamiltonian, the interacting part gives
  \be
 C= \sum_{k,k'}[(\vec{k}-\vec{k'}).\hat{n}][D(\mb{k}-\mb{k'})c^\dagger_{\mb{k}\sigma} c_{\mb{k'}\sigma}b_{\mb{k}-\mb{k'}}-H.c.].
 \ee }, we obtain:
\bea
M(z, T, T_e)&=&\frac{1}{\chi_0}\sum_{kk'} \left|D(\mb{k}-\mb{k'})\right|^2[(\vec{k}-\vec{k'}).\hat{n}]^2\nn
&&\times[{f (1-f')(1+n))-f'(1-f)n}]\nn
&&\times\frac{1}{(\epsilon_{k}-\epsilon_{k'}-\omega_{q})} \left[\frac{1}{(\epsilon_{k}-\epsilon_{k'}-\omega_{q}+z)}+ \frac{1}{(\epsilon_{k}-\epsilon_{k'}-\omega_{q}-z)}\right].\nn
\label{eq:MT}
\eea
Here, $f=f(\epsilon_k, \beta_e)$ and $f'=f(\epsilon_{k'}, \beta_e)$ are the Fermi-Dirac distribution functions at different energies such as $\epsilon_{k}$ and $\epsilon_{k'}$ and, electron temperature $T_e=\frac{1}{\beta_e}$. $n=n({\omega_{q}}, \beta)$ is the Bose-Einstein distribution function, $T=\frac{1}{\beta}$ is the phonon temperature.
$z=\w+i\delta$  and $\delta\rightarrow0^+$. Here we assume a steady-state situation in which electron temperature stays constant at $T_e$, and phonon temperature also stays constant at T. This situation can be experimentally created by a continuous laser excitation of graphene. The memory function has real and imaginary parts: $M(z,T, T_e)= M^{\prime}(\w, T, T_e)+ M^{\prime\prime}(\w, T, T_e)$. We are interested in the scattering rate which is the imaginary part of the memory function (i.e.$ M^{\prime\prime}(\w, T, T_e)= 1/\tau(\w,T, T_e)$). In that case equation (\ref{eq:MT}) can be simplified to
\bea
\frac{1} {\tau(\w, T, T_e)}& = &\frac{\pi}{\chi_0}\sum_{kk'} \left|D(\mb{k}-\mb{k'})\right|^2[(\vec{k}-\vec{k'}).\hat{n}]^2\nn
&&\times[{f (1-f')(1+n))-f'(1-f)n}]\nn
&&\times\frac{1}{\w}\Bigg[\delta(\epsilon_k-\epsilon_{k'}- \omega_{q}+\w) -\delta(\epsilon_k-\epsilon_{k'}- \omega_{q}-\w)\Bigg].
\label{eq:ImM}
\eea
 Converting the sums over momentum indices into integrals using the linear energy dispersion relation $\epsilon_k=v_Fk$ and $\epsilon_k'=v_Fk'$ and after further simplifying the above equation, we get,
\bea
  \frac{1} {\tau(\w, T, T_e)}&=&\frac{1} {\tau_0}\int_0^{q_{BG}} dq \times q^3 \sqrt{1-({q}/{2k_f})^2}\nn
&&\times\Bigg\lbrace(1-\frac{\w_q}{\w})\left[ n(\beta, \w_q)-n(\beta_e, \w_q-\w)\right]\nn && +(\rm terms\, with\, \w\,\rightarrow -\w)..\Bigg\rbrace.
\label{eq:ImMf}
\eea
 Here, $1/\tau_0=\frac{N^2 D_0^2}{32 \pi^3 \chi_0 \rho_m k_F v_s}$ and $q_{BG}$ being the Bloch-Gr\"{u}neisen momentum i.e. the maximum momentum for the phonon excitations (i.e. $v_sq_{BG} =2k_Fv_s=\Theta_{BG}$). In graphene, a new temperature crossover known as  Bloch-Gr\"{u}neisen temperature ($\Theta _{BG}$) is introduced due to small Fermi surface($k_F$) as compared to Debye surface($k_D$)\cite{LR}. Thus in this system when $k_F<<k_D$, below the Bloch-Gr\"{u}neisen temperature, only small number of phonons with wave vector ($k_{ph}<2k_F$) can take part in scattering.  Various limiting cases of equation (\ref{eq:ImMf}) are studied in the next section.
 
 \subsection{Limiting cases for the generalised Drude scattering rate}

\textbf{Case-I}: DC limit\\
 Within this limit, curly bracket in equation (\ref{eq:ImMf}) reduces to \\
 \bea
 2 \lim_{\omega \to 0} \left[ n(\beta, \w_q) - \sum_{m=0}^{\infty} \omega ^{2m}\left\lbrace\frac{\partial^{2m}}{\partial\w_q^{2m}}n(\beta_e, \w_q) + \w_q \frac{\partial^{2m+1}}{\partial\w_q^{2m+1}}n(\beta_e, \w_q)\right\rbrace \right],
 \eea
 Here we consider only m=0 i.e. the leading order case,
 \bea
  \frac{1} {\tau(\w,T, T_e)}= \frac{1}{\tau_0} \int_0^{q_{BG}} dq \times q^3\sqrt{1-\left(\frac{q}{2k_f}\right)^2}\bigg(n(\beta, \w_q) - n(\beta_e, \w_q) -\w_q n'(\beta_e, \w_q) \bigg).
 \label{eq:ImMfdc}
 \eea
Using relations $\w_q = v_s q$, $\w_{BG}\simeq\Theta_{BG} = 2v_s k_F$ and defining $\frac{\w_q}{T}=x$, $\frac{\w_q}{T_e}=y$, the above equation becomes,
\bea
  \frac{1} {\tau(\w,T, T_e)}&=&\frac{1}{\tau_0} \frac{2}{v_s^4}\bigg[T^4\int_0^{\frac{\Theta_{BG}}{T}} dx \times \frac{x^3}{e^x -1}  \sqrt{1-\left(\frac{x^2 T^2}{\Theta_{BG}^2}\right)}+ \nn
&& T_e^4\int_0^{\frac{\Theta_{BG}}{T_e}} dy \times y^3\sqrt{1-\left(\frac{y^2 T_e^2}{\Theta_{BG}^2}\right)}\nn
&&\times \bigg(\frac{y-1}{e^y -1} + \frac{y}{(e^y -1)^2}\bigg)\bigg]
\label{eq:ImMfdc1}
\eea
Subcase (a): $ T, T_e << \Theta_{BG}$, i.e., when both the phonon temperature and electron temperature are lower than the Bloch-Gr\"{u}neisen temperature. Equation (\ref{eq:ImMfdc1}) gives
\bea
 \frac{1} {\tau(T, T_e)}&=& \frac{1} {\tau_0}\frac{2}{v_s^4}\bigg[T^4\times \frac{\pi^4}{15} + T_e^4 \times \frac{\pi^4}{5})\nn
&=&\frac{1} {\tau_0}\frac{2}{v_s^4}\bigg[A_1T^4 + B_1 T_e^4\bigg].
\eea
Here $A_1= \frac{\pi^4}{15} $ and $B_1= 3A_1$.\\
Subcase (b) In high temperature case, $ T, T_e >> \Theta_{BG}$, equation (\ref{eq:ImMfdc1}) reduces to
\bea
\frac{1} {\tau(T, T_e)}&=&\frac{1} {\tau_0} \frac{2}{v_s^4}\bigg[ \frac{7}{30} T \Theta_{BG}^3 + \frac{1}{6} \Theta_{BG}^4 \bigg]\nn
&=& \frac{1} {\tau_0}\frac{2}{v_s^4}\bigg[A_2 T + B_2\bigg]
\eea
Here $A_2= \frac{7}{30}\Theta_{BG}^3 $ and $B_2= \frac{1}{6} \Theta_{BG}^4$. It is notable here that the scattering rate is independent of electron temperature, and it only depends on the phonon temperature.\\
Subcase (c) $T>>\Theta_{BG}, T_e<<\Theta_{BG}$. In this case scattering rate can be written as\\
\bea
\frac{1} {\tau(T, T_e)}&=&\frac{1} {\tau_0}\frac{2}{v_s^4}\bigg[ \frac{7}{30} T \Theta_{BG}^3 + T_e^4 \left(\frac{\pi^4}{5}\right)\bigg]\nn
&=& \frac{1} {\tau_0} \frac{2}{v_s^4}\bigg[A_3 T + B_3 T_e^4\bigg]
\eea
Here $A_3=\frac{7}{30}\Theta_{BG}^3$ and $B_3=\frac{\pi^4}{5} $. In this case $1/\tau$ leads to the linear phonon temperature dependence in high temperature regime and shows the $T_e^4$- dependence below the BG temperature.\\
Subcase (d) $T<<\Theta_{BG}, T_e>>\Theta_{BG}$. $\frac{1} {\tau(T, T_e)}$ has $T^4$- dependence. Scattering rate is independent of the electron temperature. On the other hand, when $T=T_e$, the result of scattering rate is identical as obtained in an equilibrium electron-phonon interaction in graphene case \cite{EM,LR} as expected. These results are tabulated in Table \ref{T:results}.\\
\textbf{Case-II}: Finite frequency regimes\\
Subcase (1): Consider $\w >> \w_{BG}$, then equation (\ref{eq:ImMf}) becomes
\bea
 \frac{1} {\tau(\w, T, T_e)}&=&\frac{1} {\tau_0}\int_0^{q_{BG}} dq \times q^3 \sqrt{1-({q}/{2k_f})^2}\nn
&&\times\Bigg\lbrace 2n(\beta, \w_q)-n(\beta_e, -\w)-n(\beta_e, \w) \Bigg\rbrace.
\eea
This can be simplified by setting $\frac{\w_q}{T}=x$, $\frac{\w}{T_e}=\xi$, then we have
\bea
\frac{1} {\tau(\w, T, T_e)}&=&\frac{1} {\tau_0}\frac{2}{v_s^4}T^4\int_0^{\frac{\Theta_{BG}}{T}} dx \times x^3\sqrt{1-\left(\frac{x^2 T^2}{\Theta_{BG}^2}\right)}\nn &&\left(\frac{2}{e^x -1}-\frac{1}{e^{-\xi} -1}-\frac{1}{e^\xi -1}\right).
 \label{eq:ImMfac}
\eea
After simplifying the above equation, it is observed that there is only the phonon contribution at higher frequency. To further simplify the above equation, we study the following subcases:

    In the low temperature regime $T<<\Theta_{BG}$, equation (\ref{eq:ImMfac}) becomes
    \bea
    \frac{1} {\tau(\w, T)}&=&\frac{1} {\tau_0}\frac{2}{v_s^4}T^4 \bigg[\frac{2\pi^4}{15}-\frac{1}{4}\frac{\Theta_{BG}^4}{T^4}\bigg]\nn
    &=& \frac{1} {\tau_0}\frac{1}{v_s^4}\bigg[A_5 T^4 + B_5\bigg]
    \eea
  Here, $A_5=\frac{2\pi^4}{15}$  and $B_5 = - \frac{\Theta_{BG}^4}{4}$. 
  
   In the high temperature regime $T>> \Theta_{BG}$, equation (\ref{eq:ImMfac}) takes the following form
   \bea
  \frac{1} {\tau(\w, T)}&=&\frac{1} {\tau_0}\frac{1}{v_s^4}\bigg[\frac{7}{15}\Theta_{BG}^3 T -\frac{1}{4}\frac{\Theta_{BG}^4}{T^4} \bigg]\nn
   &=&\frac{1} {\tau_0}\frac{1}{v_s^4} \bigg[A_6 T + B_5\bigg]
   \eea
   Here $A_6= \frac{7}{15}\Theta_{BG}^3$. It is also noticeable here that in both the cases $ \frac{1} {\tau(\w, T)}$ shows the  frequency independent behavior.  At $T\rightarrow 0$ and higher frequency regimes, $\frac{1} {\tau(\w)}$ shows saturation.\\
   Subcase (2): At finite but lower frequency $\w << \w_{BG}$ case, with relation $\w_q=v_s q$ the equation (\ref{eq:ImMf}) becomes
  \bea
 \frac{1} {\tau(\w, T, T_e)}&=&\frac{1} {\tau_0}\int_0^{\Theta_{BG}} dq \times \w_q^3 \sqrt{1-(\frac{\w_q}{\Theta_{BG}})^2}\bigg[\frac{1}{e^\frac{w_q}{T}-1}-\nn
 &&\sum_{m=0}^{\infty} \omega ^{2m}\left(\frac{\partial^{2m}}{\partial\w_q^{2m}}\frac{1}{e^\frac{w_q}{T_e}-1} + \w_q \frac{\partial^{2m+1}}{\partial\w_q^{2m+1}}\frac{1}{e^\frac{w_q}{T_e}-1}\right)\bigg]
 \label{eq:ImMfacl}
  \eea
 This is the general equation of the imaginary part of memory function when frequency is lower than the Bloch-Gr\"{u}neisen frequency. The above equation can be further simplified by setting the variables $\frac{\w_q}{T}=x$, $\frac{\w_q}{T_e}=y$, and for m=1,  the equation (\ref{eq:ImMfacl}) reduces 
 \bea
 \frac{1} {\tau(\w, T, T_e)}&=&\frac{1} {\tau_0}\frac{2}{v_s^4}\bigg[T^4\int_0^{\frac{\Theta_{BG}}{T}} dx \times x^3\sqrt{1-\left(\frac{x^2 T^2}{\Theta_{BG}^2}\right)}\frac{1}{e^x -1} + \nn
&& \w^2T_e^2\int_0^{\frac{\Theta_{BG}}{T_e}} dy \times y^3\sqrt{1-\left(\frac{y^2 T_e^2}{\Theta_{BG}^2}\right)}\nn
&&\times\left(n_y + 3n_y^2 +2n_y^3 - y \left(n_y -7n_y^2 -12n_y^3-6n_y^4\right)\right)\bigg]
 \eea   
Here, $n_y =\frac{1}{e^y -1}$.  Further we study the frequency dependent scattering rate at low and high temperature regimes of both electron and phonon sub-systems. We consider first two terms (m=0 and m=1) in the series of the equation (\ref{eq:ImMfacl}). The analytic results obtained in the present subcase ($\w << \w_{BG}$) are presented in Table \ref{T:results}. It is observed that there is $\w^2$-dependence multiplied by the electron temperature in the lower frequency regime. In the general case, numerical computations of equation (\ref{eq:ImMf}) is presented in the next section. And in the appropriate limiting cases, numerical results agree with analytical results presented in Table \ref{T:results}.
    
\begin{center}
\begin{table}[h]
\begin{tabular}{|l|l|l|}
\hline
	No & Regimes & $ \frac{1} {\tau}$ \\
\hline
 \hline
	1 & $\w= 0;\,\,\,\,T_e, T<<\Theta_{BG}$ & $A_1T^4 + B_1 T_e^4.$\\
\hline
	 & $\w= 0;\,\,\,\, T_e, T>>\Theta_{BG}$  & $A_2T+ B_2$ \\
\hline
	 & $\w= 0;\,\,\,\, T>>\Theta_{BG}, T_e<<\Theta_{BG}$  & $A_3T+B_3T_e^4$. \\
\hline
	 & $\w= 0;\,\,\,\, T<<\Theta_{BG},T_e>>\Theta_{BG}$  & $A_4T^4$+ constant. \\
\hline
2  &  $\w >> \w_{BG};\,\,\,\, T<<\Theta_{BG}$ & $A_5T^4$.\\
\hline
   & $\w >> \w_{BG};\,\,\,\, T>>\Theta_{BG}$ & $A_6T$. \\
 \hline
 3 & $\w << \w_{BG};\,\,\,\, T, T_e >> \Theta_{BG}$ & $A_7T + B_7 T_e + C_7 \w^2 T_e$.\\
 \hline
   & $\w << \w_{BG};\,\,\,\, T, T_e << \Theta_{BG}$ & $A_8T^4 + B_8 T_e^4 + C_8 \w^2 T_e^2$. \\
   \hline
    & $\w << \w_{BG};\,\,\,\, T>>\Theta_{BG}, T_e << \Theta_{BG}$ & $A_9T + B_9 T_e^4 + C_9 \w^2 T_e^2$.\\
    \hline
    & $\w << \w_{BG};\,\,\,\, T<<\Theta_{BG}, T_e >>\Theta_{BG}$ & $A_{10}T^4 + B_{10} T_e   + C_{10} \w^2 T_e $.\\
 \hline

\end{tabular}
\caption{The results of electrical scattering rate due to the electron-phonon interactions in different limiting cases. Here, $A_7=\frac{7}{30}\Theta_{BG}^3$, $B_7=-\frac{7}{30}\Theta_{BG}^3$,$C_7 =\frac{20}{3}\Theta_{BG}$, and  $A_8=\frac{\pi^4}{15}$, $B_8 = \frac{\pi^4}{5}$, $C_8=5\pi^2 -\zeta(5)$ and $A_9=\frac{7}{30}\Theta_{BG}^3$, $B_9=\frac{\pi^4}{5}$, $C_9=C_8 = constant$, and $A_{10} =\frac{\pi^4}{15}$, $B_{10} = -\frac{7}{30}\Theta_{BG}^3$ and $C_{10} = \frac{20}{3}\Theta_{BG}$.}
 \label{T:results}
 \end{table}
\end{center}



\section{Numerical analysis} 
\label{sec:numeric}

\begin{figure}[h]
	\begin{subfigure}{0.48\textwidth}
		\includegraphics[width=1\linewidth]{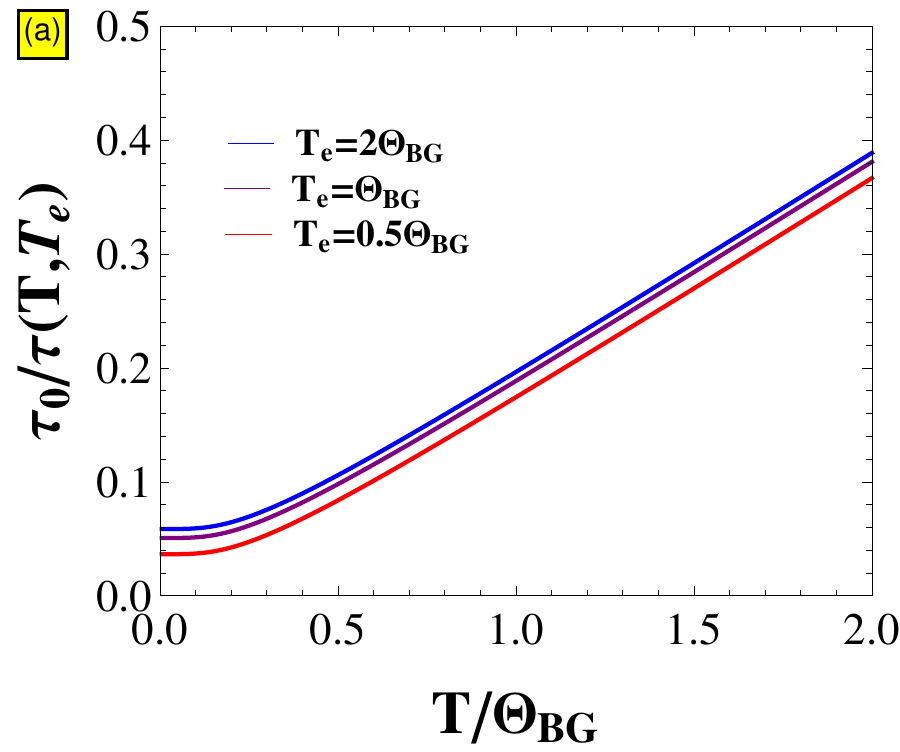}
		\caption{}\label{fig:a}
	\end{subfigure}
	\hspace*{\fill}  
	\begin{subfigure}{0.48\textwidth}
		\includegraphics[width=1\linewidth]{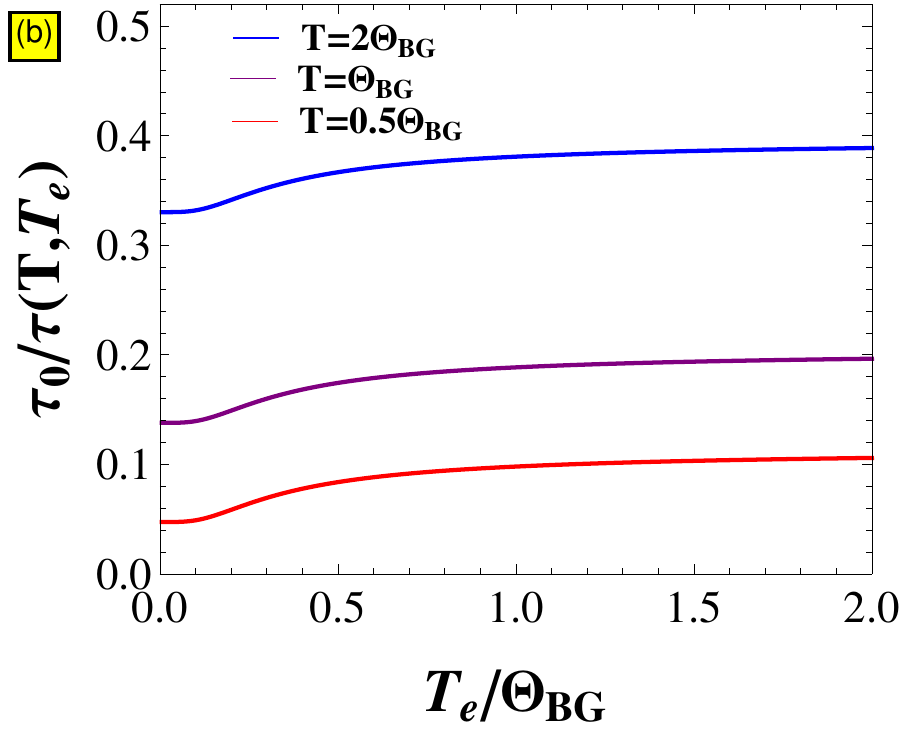}%
		\caption{}\label{fig:b}
	\end{subfigure}
	
	\bigskip  
	
	\begin{subfigure}{0.48\textwidth}
		\includegraphics[width=1\linewidth] {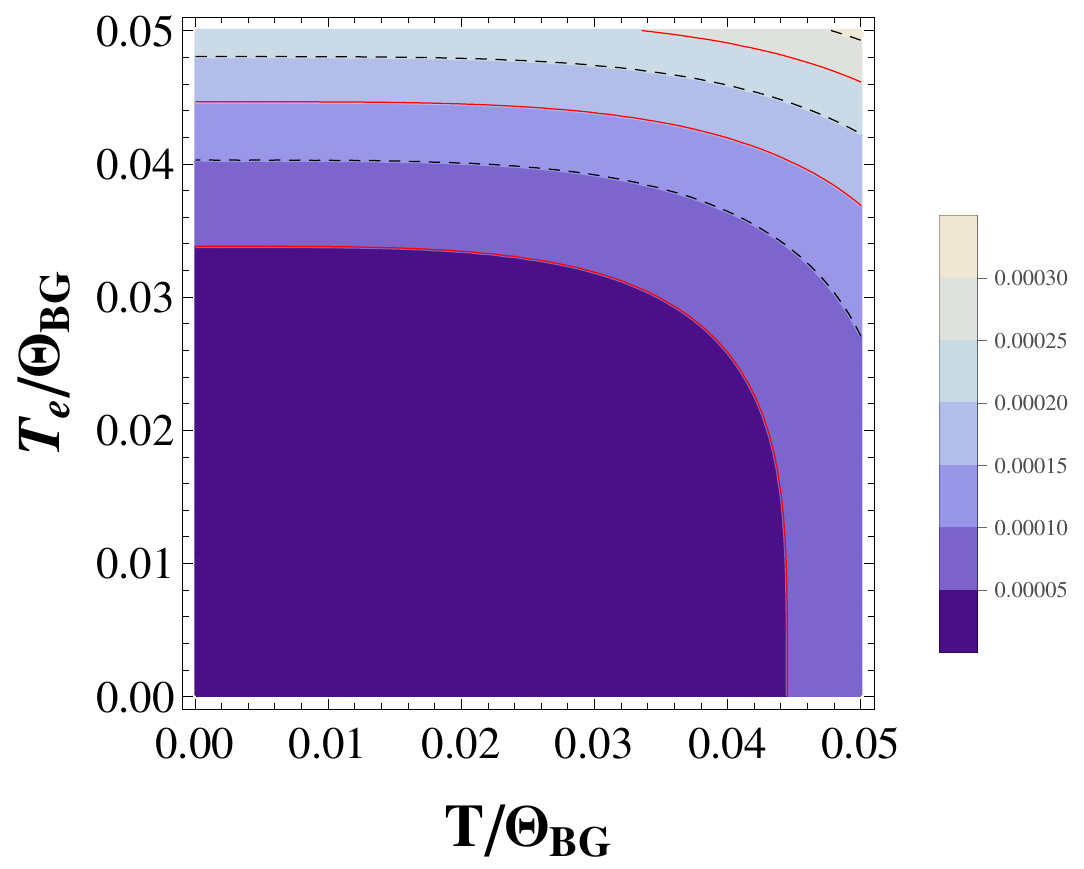}
		\caption{}\label{fig:c}
	\end{subfigure}
	\hspace*{\fill} 
	\begin{subfigure}{0.48\textwidth}
		\includegraphics[width=1\linewidth] {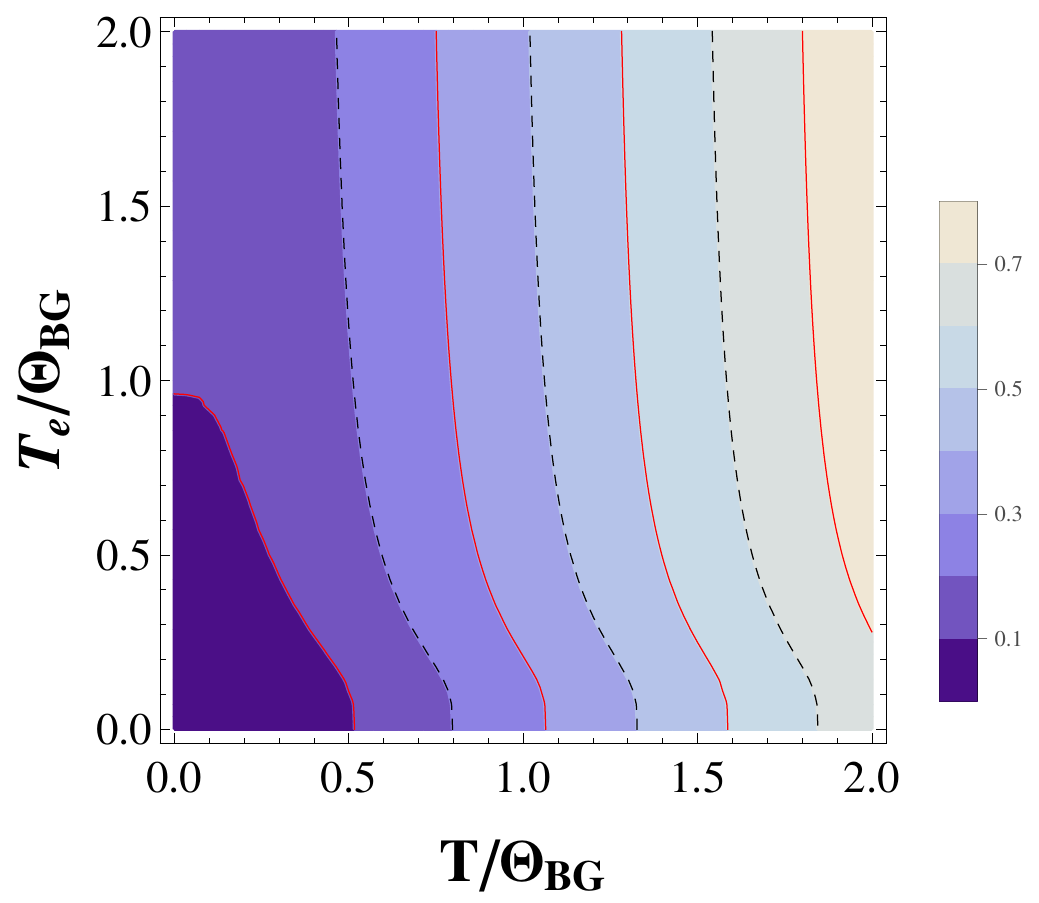}%
		\caption{}\label{fig:d}
	\end{subfigure}
	\centering
	 \caption{(a) Variation of the scattering rate with phonon temperature at zero frequency and different electron temperatures. (b) Variation of the scattering rate with electron temperature at zero frequency and different phonon temperatures. Here both the electron and phonon temperatures are scaled with the Bloch-Gr\"{u}neisen temperature and $1/\tau(T, T_e)$ is scaled with $1/\tau_0$. Figures (c) and (d) depict contour plots T vs $T_e $ for the scattering rate at zero frequency.}
	 	\label{fig:fig1}
\end{figure}

We have numerically computed the equation (\ref{eq:ImMf}) in different frequency and temperature regimes. In Fig.\ref{fig:fig1}(a), we depict the phonon temperature dependence of scattering rate $1/\tau(T, T_e)$ normalized by $1/\tau_0( = \frac{N^2 D_0^2}{16 \pi^3 \chi_0 \rho_m k_F v_s^5})$ at zero frequency and at different electron temperatures. From Fig.\ref{fig:fig1}(a), we observe that at high temperatures ($T_e, T>>\Theta_{BG}$), $1/\tau(T, T_e)\propto T$. This can also be seen in the corresponding case ($T_e, T>>\Theta_{BG}$) in Table \ref{T:results}.
 At very low temperature ($ T, T_e << \Theta_{BG}$), $1/\tau\propto T^4 \, and \,  T_e^4$.
 Fig.\ref{fig:fig1}(b) shows the dependence of $1/\tau$ on $T_e$ in the DC limit. It is observed that $1/\tau$  is independent of $T_e$ when $T_e >> \Theta_{BG}$.
Contour plots (Fig.\ref{fig:fig1}(c) and Fig.\ref{fig:fig1}(d)) depict the constant value of $1/\tau$ in $T_e$ and T plane. The contour for higher values of T and $T_e$ are for higher $1/\tau$. 

 From the contour plots, we notice that they are not symmetric around $T=T_e$ line. The physical reason for this asymmetry is that the scattering rate is differently effected by phonon temperature and electron temperature (the pre-factor $A_1$ of $T^4$ term is not equal to the prefactor $B_1$ of $T_e^4$ terms). At very low temperature $T^4$ behavior is due to Pauli blocking effect. We notice that at high temperature ($T_e, T>>\Theta_{BG}$), $1/\tau(T, T_e)$ is proportional to $T$, not $T_e$. The reason for this behavior is that at high temperatures phonon modes scale as $k_B T (<n_q>=\frac{1}{e^{\beta \w_q} -1} \propto k_B T)$, thus scattering increases with increasing temperatures linearly. For $T_e >>\Theta_{BG} $ the electron distribution can be approximated as Boltzmann distributions because $\Theta_{BG} \simeq T _F$ (the Fermi temperature). The temperature effect is exponentially reduced in this case as compared to phonons ($<n_q> \propto T$). Thus at high temperatures, the scattering rate is proportional to T.

\begin{figure}[h]
	\begin{subfigure}{0.48\textwidth}
		\includegraphics[width=1\linewidth]{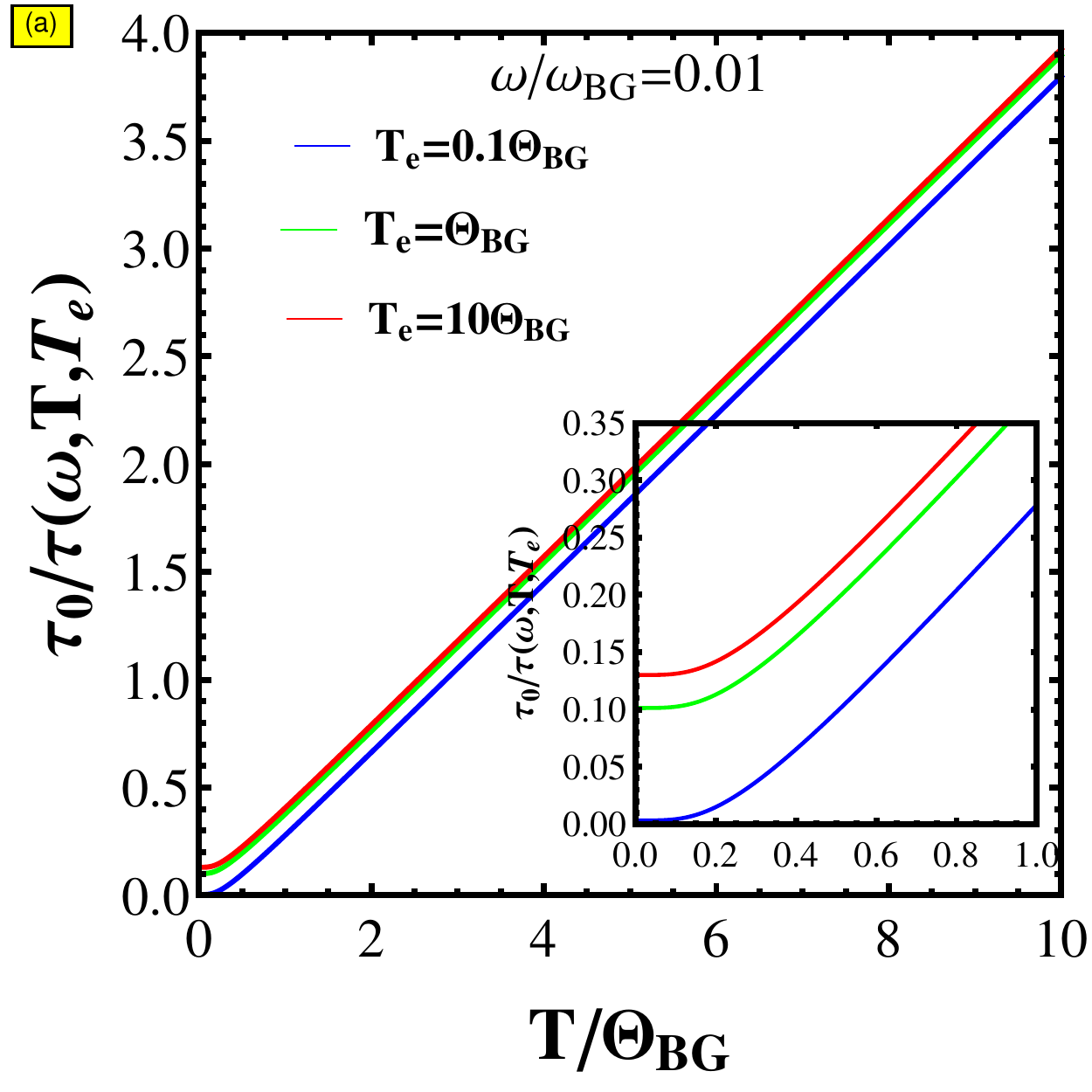}
		\caption{}\label{fig:a}
	\end{subfigure}
	\hspace*{\fill}  
	\begin{subfigure}{0.48\textwidth}
		\includegraphics[width=1\linewidth]{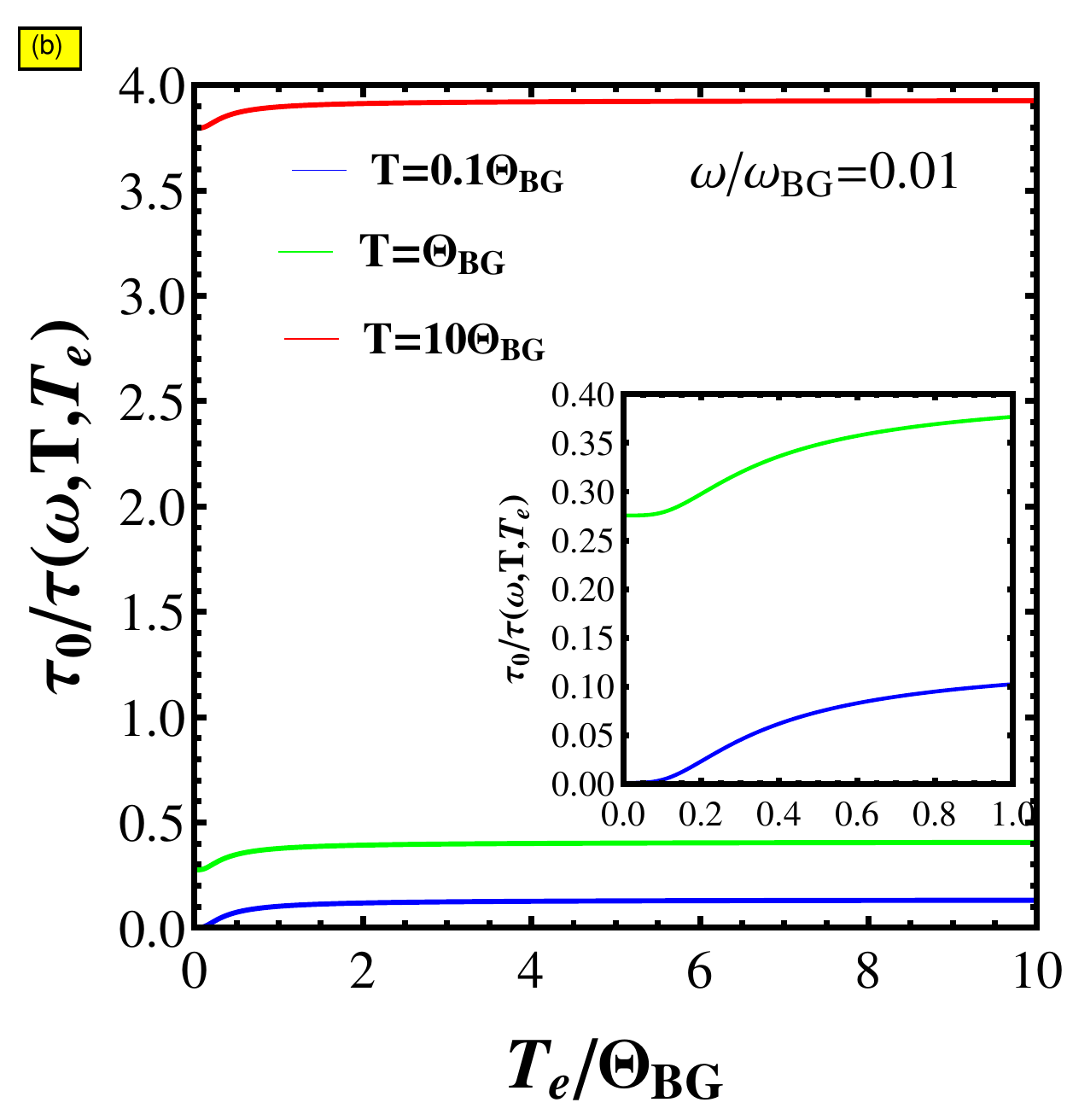}%
		\caption{}\label{fig:b}
	\end{subfigure}
 \caption{(a)Variation of the scattering rate with phonon temperature at finite but lower frequency and at different electron temperatures, and inset shows the lower phonon temperature range. (b)Variation of the scattering rate with electron temperature at finite frequency and at different phonon temperatures, and inset shows the lower electron temperature range.}
 \label{fig:fig2}
\end{figure}

In Fig.\ref{fig:fig2}(a), we plot the phonon temperature dependency of scattering rate $\tau_0/\tau(\w,T, T_e)$ at lower frequency and at different temperatures of electrons. It is observed that at lower phonon temperature range, the magnitude of scattering rate increases with increasing temperature as $T^4$ behavior. At higher T it shows T-linear behavior.
In  Fig.\ref{fig:fig2}(b), the variation of electron temperature dependence of $\tau_0/\tau(\w,T, T_e)$ at different phonon temperature scaled with BG temperature is shown. The insets of both the figures show low temperature behavior ($T, T_e<<\Theta_{BG}$). The low frequency behavior is similar to the DC case.

\begin{figure}[h]
	\begin{subfigure}{0.46\textwidth}
		\includegraphics[width=1\linewidth]{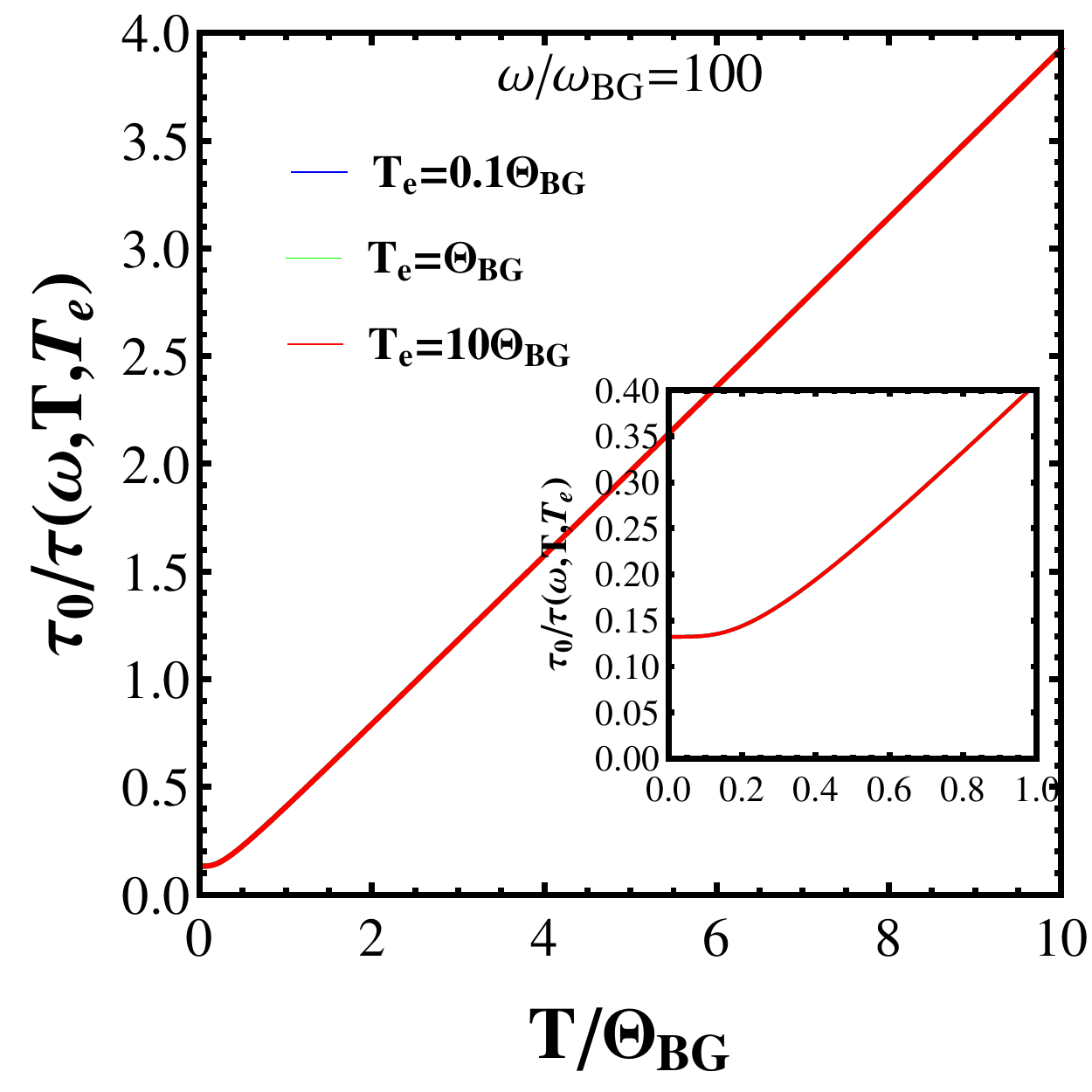}
		\caption{}\label{fig:a}
	\end{subfigure}
	\hspace*{\fill}  
	\begin{subfigure}{0.50\textwidth}
		\includegraphics[width=1\linewidth]{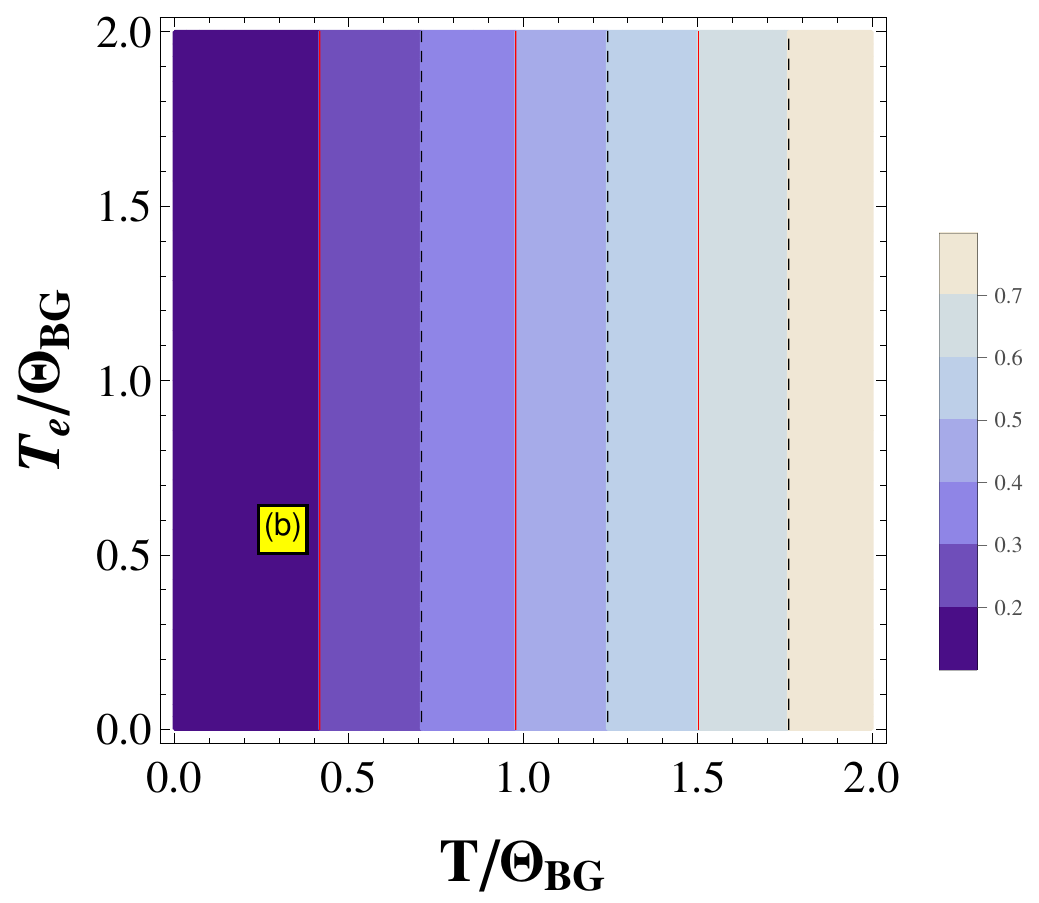}%
		\caption{}\label{fig:b}
	\end{subfigure}
 \caption{(a)Variation of the scattering rate with phonon temperature at higher frequency ($\w /\w_{BG}=100$) and at different electron temperatures, and inset shows the same plot with lower phonon temperature range over $\Theta_{BG}$. The inset also shows finite scattering rate even at zero phonon temperature which is nothing but the non-equilibrium generalization of the Holstein mechanism \cite{Das,Singh, LR}. (b) The contour plot depicts the behavior of scattering rate at higher frequency ($\w /\w_{BG}=100$).}
           \label{fig:fig3}
	\end{figure}

 In order to study the higher frequency regime, we plot the variation of the scattering rate with phonon temperature at higher frequency ($\w /\w_{BG}=100$) and at different electron temperatures in Fig.\ref{fig:fig3}(a). It is observed that at higher frequency, scattering rate is independent of the electron temperature (compare with the corresponding entry given in Table \ref{T:results}). Plot shows the T-linear behavior above BG temperature and $T^4$ behavior below lower BG temperature. These results agree with the result of \cite{Efetov,EH}. At higher frequency, the scattering rate is controlled by phonon temperature. The independence of $1/\tau$ from $T_e$ is also shown in the contour plot (Fig.\ref{fig:fig3}(b)).

\begin{figure}[htbp!]
        \centering
        \includegraphics[angle=0,width=6cm]{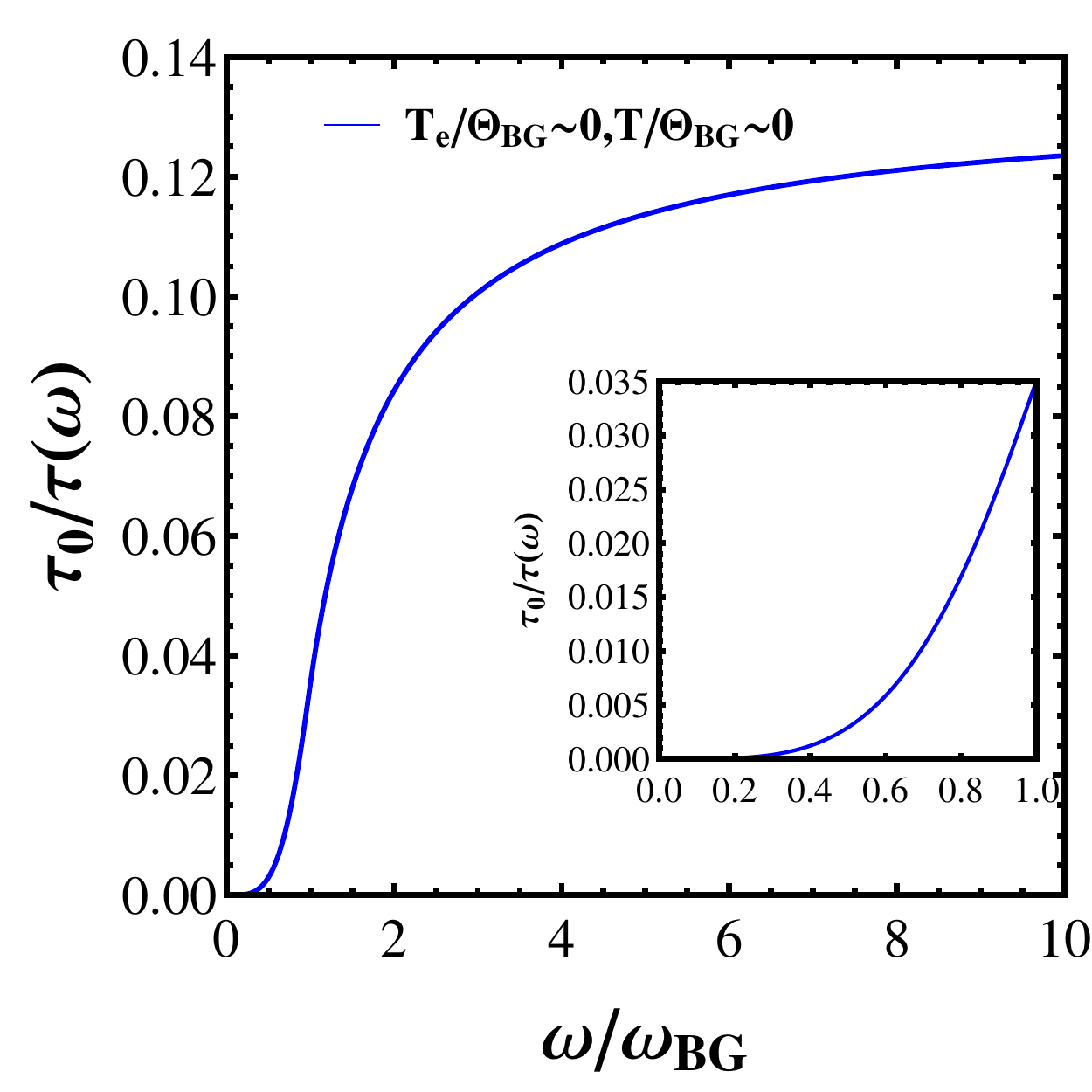}
           \caption{(a)Variation of the scattering rate with frequency at zero electron and phonon temperatures, and inset shows the same plot at lower frequency over Bloch-Gr\"{u}neisen frequency.}
           \label{fig:fig4}
	\end{figure}

 We further analyzed the scattering rate at zero temperature in which both electron subsystem and phonon subsystem are at zero temperature. In this regime $1/\tau$ scales as $\omega^4$ as depicted in Fig.\ref{fig:fig4}.


	\begin{figure}[htbp!]
        \centering
        \includegraphics[angle=0,width=7.2cm]{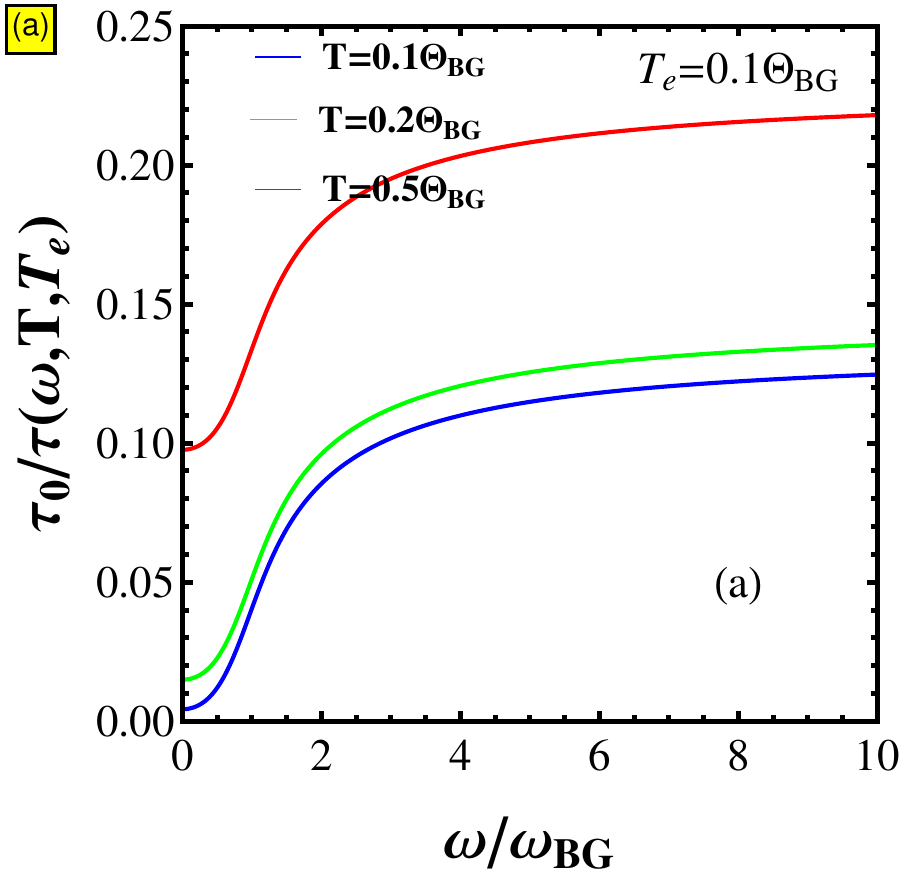}
        \includegraphics[angle=0,width=7.2cm]{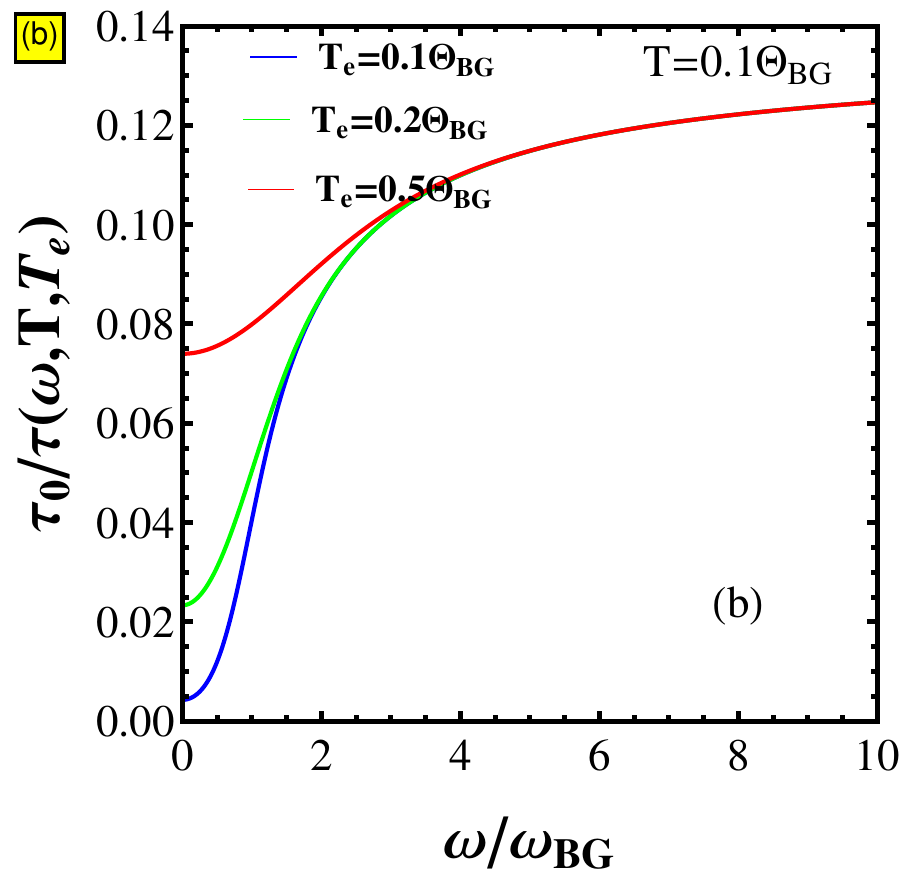}
           \caption{Variation of the scattering rate with frequency at different electron and phonon temperatures.}
           \label{fig:fig5}
	\end{figure}	

	To order to study the scattering rate with frequency, we plot the frequency dependence behavior of the scattering rate $1/\tau(\w,T, T_e)$ at different temperatures of electron and phonon subsystems  in Fig.\ref{fig:fig5}.  Fig.\ref{fig:fig5}(a) depicts the variation of scattering rate with frequency at different phonon temperatures and at fixed electron temperature. At higher frequency, $1/\tau$ saturates and at lower frequency it shows $\omega^2$ behavior.

	In Fig.\ref{fig:fig5}(b), we plot the variation of scattering rate with frequency at different electron temperatures and fixed phonon temperature. From Fig.\ref{fig:fig5}(b), it is clear that on increasing the electron temperature, scattering rate increases in lower frequency regime but scattering rate goes into saturation trend in the high frequency regimes, and become independent of electron temperature. This can also  be obtained from Table \ref{T:results} (in the $\omega>>\omega_{BG}$ case).

\section{Conclusion and discussion}	
\label{sec:conclusion}

We presented a theoretical study of non-equilibrium relaxation of electrons due to their coupling with phonons in graphene by using the memory function approach.
In our results at zero frequency limit, it is observed that if both the electron and phonon temperature are not same, DC scattering rate has a fourth power law  behavior of both the electron and phonon temperaures i.e. ($A_1T^4 + B_1 T_e^4$) below the BG temperature. While at higher temperature, $1/\tau$ shows the T-linear dependency only (it does not depend on $T_e$). Further, it is important to notice here that DC scattering rate and AC scattering rate shows the similar T-linear behavior at higher temperature. 
\begin{center}
	\begin{table}[h]
		\begin{tabular}{|l|l|l|l|}
			\hline
			No & Regimes & Graphene $\bigg (\dfrac{1}{\tau} \bigg)$ &   Metals $\bigg(\dfrac{1}{\tau}\bigg)$\cite{Das}\\
			&     & 2D   & 3D\\
			&     & Bloch Gr\"unisen & Debye \\
			&     &Temperature $(\Theta_{BG})$  & Temperature $(\Theta_D)$\\
			\hline
			\hline
			1 & $\w= 0;\,\,\,\,T_e, T<<\Theta_{BG}, \Theta_D$ & $A_1T^4 + B_1 T_e^4.$ & $a_1T^5+b_1T_e^5.$\\
			\hline
			& $\w= 0;\,\,\,\, T_e, T>>\Theta_{BG}, \Theta_D$  & $A_2T$. & $a_2+b_2T$\\
			\hline
			& $\w= 0;\,\,\,\, T>>\Theta_{BG}, T_e<<\Theta_{BG}$  & $A_3T+B_3T_e^4$.& -\\
			\hline
			& $\w= 0;\,\,\,\, T<<\Theta_{BG},T_e>>\Theta_{BG}$  & $A_4T^4$. &-\\
			\hline
			2  &  $\w >> \w_{BG},\w_{D};\,\,\,\, T>>\Theta_{BG}, \Theta_D$ & $A_5T$.& $a_3+b_3T$.\\
			\hline
			& $\w >> \w_{BG},\w_D;\,\,\,\, T<<\Theta_{BG}, \Theta_D$ & $A_6T^4$.& $a_4 + b_4T^5$ \\
			\hline
			3 & $\w << \w_{BG},\w_{D};\,\,\,\, T, T_e >> \Theta_{BG}, \Theta_D$ & $A_7T + B_7 T_e + C_7 \w^2 T_e$.& $a_5T +b_5\w^2 T_e$\\
			\hline
			& $\w << \w_{BG},\w_{D};\,\,\,\, T, T_e << \Theta_{BG},\Theta_D$ & $A_8T^4 + B_8 T_e^4 + C_8 \w^2 T_e^2$.& $a_6T^5+b_6T_e^5+c_6T_e^5\w^2$ \\
			\hline
			& $\w << \w_{BG};\,\,\,\, T>>\Theta_{BG}, T_e << \Theta_{BG}$ & $A_9T + B_9 T_e^4 + C_9 \w^2 T_e^2$.&-\\
			\hline
			& $\w << \w_{BG};\,\,\,\, T<<\Theta_{BG}, T_e >>\Theta_{BG}$ & $A_{10}T^4 + B_{10} T_e   + C_{10} \w^2 T_e$.&-\\
			\hline
		\end{tabular}
		\caption{Comparison of non-equilibrium electron relaxation in metals and in graphene}
		\label{T:result 2}
	\end{table}
\end{center}
 
In Table \ref{T:result 2}, we compare the results of scattering rates for the simple metals and the present case of graphene. We observed that $T^5$-law of $1/\tau$ in the case of metals (in regimes $ \omega=0,\, T <<\Theta_D$) changes to $T^4$-law in the corresponding case in graphene. However, in the case of high temperatures and high frequencies, temperature dependence of $1/ \tau$ in both metals and in graphene remains the same.

At higher frequency, the scattering rate is controlled by phonon temperature in both the cases (of metals and graphene). In the low frequency case ($\omega<<\omega_{D}$) and in lower temperature regimes ($T, T_e<<\Theta_D$) $1/\tau$ in metals has three terms ($a_6T^5+b_6T_e^5+c_6T_e^5\w^2$) whereas in the corresponding case of graphene this dependence changes to ($A_8T^4 + B_8 T_e^4 + C_8 \w^2 T_e^2$). These results can be verified that in a typical pump-probe experiments \cite{Shah,Verburg,Liaros}.


\end{document}